# Deep Anatomical Federated Network (Dafne): An open client-server framework for the continuous, collaborative improvement of deep learning-based medical image segmentation

Francesco Santini, PhD[1,2*], Jakob Wasserthal, PhD[2], Abramo Agosti, PhD[3], Xeni Deligianni, PhD[1,2], Kevin R. Keene, PhD[4], Hermien E. Kan, PhD[5], Stefan Sommer, PhD[6,7,8], Fengdan Wang, MD[9], Claudia Weidensteiner, PhD[1,2], Giulia Manco, MSc[10], Matteo Paoletti, MD[11], Valentina Mazzoli, PhD[12], Arjun Desai, PhD[13,14], and Anna Pichiecchio, MD[11,15]

1. Basel Muscle MRI, Department of Biomedical Engineering, University of Basel, Basel, Switzerland
2. Department of Radiology, University Hospital Basel, Basel, Switzerland
3. Department of Mathematics, University of Pavia, Pavia, Italy
4. Department of Neurology, Leiden University Medical Center, Leiden, The Netherlands
5. C.J. Gorter MRI Center, Department of Radiology, Leiden University Medical Center, Leiden, The Netherlands
6. Siemens Healthineers International AG, Zurich, Switzerland
7. Swiss Center for Musculoskeletal Imaging (SCMI), Balgrist Campus, Zurich, Switzerland
8. Advanced Clinical Imaging Technology (ACIT), Siemens Healthineers International AG, Lausanne, Switzerland
9. Peking Union Medical College Hospital, Beijing, China
10. Istituti Clinici Scientifici Maugeri IRCCS, Servizio di Diagnostica per Immagini - Istituto di Montescano, Italy.
11. Advanced imaging and radiomics center, Neuroradiology Department, IRCCS Mondino Foundation, Pavia, Pavia, Italy
12. Department of Radiology, NYU Grossman School of Medicine, New York, NY, USA
13. Department of Radiology, Stanford University, Stanford, CA, USA
14. Department of Electrical Engineering, Stanford University, Stanford CA, USA
15. Department of Brain and Behavioural Sciences, University of Pavia, Pavia, Italy

Article type: Original Research, Word count: 3323

Funding information: This work has been partly supported by the Swiss National Science Foundation (SNF) Spark grant number CRSK-3_196515

Data sharing statement: The data used to produce the images of this manuscript, and the analysis workflow is available under CC-BY license at https://github.com/dafne-imaging/dafne-evaluation

**Correspondence/Originating institution:**
Francesco Santini
Department of Radiology
University Hospital Basel,
Petersgraben 4
4031 Basel
Switzerland
Tel: +41 61 55 65417

**Key points:**

- Dafne (deep anatomical federated network) is a deep learning system for medical image segmentation that implements a privacy-preserving client and server architecture, where the client refines model predictions and the model continuously evolves with distributed usage.
- Dafne was validated locally on 38 MRI datasets of the lower legs, showing ability to improve and adapt to new contrasts.
- Real-world performance is also demonstrated by showing the distributed performance over more than 600 datasets analyzed by institutional users worldwide.

# Abstract

Purpose: To present and evaluate Dafne (deep anatomical federated network), a freely available decentralized, collaborative deep learning system for the semantic segmentation of radiological images through federated incremental learning.

Materials and Methods: Dafne is free software with a client-server architecture. The client side is an advanced user interface that applies the deep learning models stored on the server to the user's data and allows the user to check and refine the prediction. Incremental learning is then performed at the client's side and sent back to the server, where it is integrated into the root model. Dafne was evaluated locally, by assessing the performance gain across model generations on 38 MRI datasets of the lower legs, and through the analysis of real-world usage statistics ($n = 639$ use-cases).

Results: Dafne demonstrated a statistically improvement in the accuracy of semantic segmentation over time (average increase of the Dice Similarity Coefficient by 0.007 points/generation on the local validation set, $p < 0.001$). Qualitatively, the models showed enhanced performance on various radiologic image types, including those not present in the initial training sets, indicating good model generalizability.

Conclusion: Dafne showed improvement in segmentation quality over time, demonstrating potential for learning and generalization.

# Main body

## Introduction

Semantic segmentation of medical images is a challenging yet important step towards the automation of both research and clinical workflows. Such segmentation supports radiologists in highlighting the presence of lesions and serves as a necessary step in extracting quantitative biomarkers, such as organ size and volume, and average tissue properties over anatomical regions.

The current highest-performing methods for automated image segmentations are based on deep neural networks, typically trained on manually labeled data (1–3). One of the key factors that contribute to the success of these algorithms is the availability of large and diverse training datasets. A high volume of training data allows the algorithm to learn a wide range of features and patterns, which enables it to generalize better to new and unseen examples (4). High variability in the training data helps the algorithm to learn robust and invariant representations, allowing it to perform well in various conditions and environments (5).

A now common solution to the difficulty of collecting all the necessary training data in a single institution is the use of federated learning (6), or, in its decentralized form, swarm learning (7,8). In federated learning, the same model is trained at different locations on local data, and the model parameters are then aggregated and redistributed multiple times for various training cycles, with the goal of obtaining a final, stable model. This approach is highly appealing in healthcare, as it largely overcomes the legal and practical hurdles of sharing patient data (9). Indeed, a number of existing studies have been focused on lesion detection and/or classification (10–14), finding similar performance between centralized and federated learning (15).

By leveraging federated learning, we propose a new pragmatic approach in medical image segmentation, alongside the traditional training-validation-deployment model where the model is optimized for a specific data type, towards a practical workflow that exploits human knowledge and the collaborative effort of the community to continuously refine segmentation models.

In this paper, we present an open-source, multiplatform client-server software system that integrates deep learning segmentation models with an advanced user interface for computer-assisted manual segmentation tasks. While the interface already implements several advanced features that make it a potentially useful tool on its own, the strength of the system lies in the seamless coupling of this interface with an incremental upgrade of the model performed on the client's side.

This system was designed with the following fundamental principles in mind, reflecting the features stated above:

16. The program has an intuitive user interface with advanced manual editing features (user convenience).

17. Data are only handled by the client and never transmitted (privacy preservation).
18. Incremental learning is performed on the client's side (resource sparing).
19. The automated segmentation is always checked and refined by the operator (waiver of medical responsibility).

The system was named Dafne (an acronym for Deep Anatomical Federated Network), and currently publicly offers four models (thigh, lower leg, abdomen, and lumbar spine), in addition to a tool to automatically train new models. Of these models, the thigh and lower leg models were the first ones to be released and were tested in this study.

# Materials and Methods

## System architecture and workflow

Dafne implements a client-server architecture. The server, including a website containing the documentation, is hosted on a Google Cloud Virtual Machine (Google LLC, Mountain View, CA, USA) at the web address https://dafne.network/. The role of the server is to provide the available segmentation models to the client and receive updated versions from the users and integrate them into the central model.

The Dafne client is a complete computer assisted segmentation platform (Fig. 1), available as Free and Open Source Software (FOSS). A personal API key, currently also distributed for free through a request form on the website, enables communication between the client and the server by means of a representational state transfer (REST) Application Programming Interface (API). Both the client, the server, and the supporting libraries are available as Free and Open Source Software at https://github.com/dafne-imaging/.

The typical Dafne workflow is shown in Fig. 2. With the segmentation task, the latest model for the selected body region is downloaded from the server and applied to one or more slices. After segmentation, the user checks and refines the proposed regions with the provided user interface tools. Both a "mask" editing mode, where the region of interest is painted with a brush-like cursor, and a "contour" editing mode, where the handles of an interpolating spline are used to define the regions of interest, are available. Refinement is made convenient by a set of advanced editing features, including nonrigid-registration-based area propagation and interpolation, image-gradient-based contour snapping, and common mask operations (area opening and closing, thresholding). A full description of the features is available in the user documentation (16).

Once the user has checked and refined the automatic segmentation, an incremental learning step is performed on the new data for a fixed duration of 5 epochs.

The refined model is then automatically sent back to the server, which has exemplary data with the corresponding "reference standard" segmentation stored privately. The received model is first validated on these private data, and then, if successful, merged with the previous version of the model, validated again, and made available for the next use. Validation is performed by calculating the Dice Similarity Coefficient (DSC) (17) between the stored validation masks and the output of the new model. If this coefficient is above a predefined level, specified in the server configuration file and currently set to 0.7, the model is made available for the next user.

## Model details

Implementation of the deep learning model is designed to be generic and extensible through a plugin-like architecture. In our implementation, the models are represented by files that contain a serialized (using the dill python package (18)) representation of a Python object, implementing the functionality of a segmentation model in the most generic way possible. This object not only contains the weights of the model, but also the functions that perform the initialization, the pre- and post-processing of the data, the incremental learning, and the averaging of the model weights for the federation step. This approach allows the implementation of virtually arbitrary segmentation algorithms and is potentially agnostic to the underlying machine learning framework.

The models currently provided by the Dafne server share the same basic implementation, based on the previous work by Agosti et al (19). The network architecture is based on the VNet (3) and ResNet (20) convolutional networks for 2D image segmentation, employing a standard encoder/decoder architecture with long and short skip connections to improve the convergence. The models were initially pretrained on 44 proton-density-weighted MRI datasets (with dataset referring to the images from a single scan of a patient) in order to provide an initial working implementation for users and to avoid catastrophic forgetting (21). The network architecture, training data and training procedure are described in detail in Agosti et al (19), and implemented using the Tensorflow (22) and Keras (23) libraries. The same models were adapted into Dafne to support single- and two-sided limb segmentation by using the same model weights and adapting the preprocessing to the different situations.

Incremental learning is performed on each used model every time a segmentation on a minimum of five slices is performed by a user. Each model undergoes five training epochs on the refined images using a batch size of 5 and an Adam optimizer (AMSGrad variant) (24,25). A class-balanced weighted cross-entropy loss function is employed, where the weights are calculated in a preprocessing step and compensate the different frequency of pixels for each segmentation class, while focalizing the attention of the network between neighboring muscles, as in (19).

## Local validation

We systematically tested the performance of the proposed approach with a retrospective analysis of anonymized clinically collected data performed under a waiver of the requirement for informed consent by the local ethical committee. It was performed by analyzing 38 T1-weighted MRI datasets of patients with suspected myositis who received a routine MRI examination including the lower leg. Deidentification was performed through an anonymized PACS search between June 2019 and June 2022 (3 years) using specialized software (26) and removal of the identifying DICOM parameters. The “Leg” model was tested. This image contrast type was not part of the original datasets used for pretraining of the models.

The 38 datasets were split into two groups, group A containing 25 datasets, and group B containing the remaining 13 (2:1 split). Each group was randomly split between two independent annotators (FS, with more than 8 years of experience in muscle MRI, and JW, with more than 3 years of experience in image analysis), working on different workstations, to mimic the real-world situation of distributed usage.

At least 5 slices in the leg region of each dataset in group A were segmented by the two annotators using the Dafne standard workflow. This means that the datasets in group A were progressively included in the training for the later datasets. The average Dice Similarity Coefficients (DSCs) of all the segmented muscles were used to evaluate the quality of the model.

After the segmentation of group A, the same annotators used the Dafne workflow to similarly segment the datasets of group B. However, the model updates generated by group B were excluded from subsequent evaluation. This division ensured that the evaluated model generations were never trained on any data belonging to group B.

Model evolution was evaluated by applying each model generation derived from segmentation of group A datasets to all the datasets in groups A and B and calculating the DSC of the automatically generated maps with the manually segmented ones. The difference in DSC between the first generation and each subsequent generation was considered (differential DSC).

A linear mixed effects model with random slope (dependent on the dataset), with model generation as the independent variable, was fitted to the time series to evaluate whether the models were able to improve their performance under this proposed workflow.

The statistical analysis was performed using Python version 3.10 with the statsmodels package version 0.13.2.

For complete reproducibility, the relevant data and statistical analysis is publicly available online under a CC-BY license (27).

## Usage statistics

User data statistics in the form of average DSC across all regions of interest and all slices were automatically sent by the client to the server and recorded every time a user finished the segmentation of a dataset. The users were not instructed on a particular segmentation style (e.g. how much margin to leave between adjacent muscles), nor were they restricted to the type of protocol, contrast, or pathology that they could import into the system, as the hypotheses were that the model could generalize to a wide range of inputs and that it would converge towards a consensus segmentation style. No patient or pathology data were collected from the sites.

The user statistics were observed over a period spanning from July 1st, 2021, to December 31st, 2022, and consisted of the collection of averaged DSC for every segmented dataset. The log file also showed how many unique users utilized the system. Users that were registered to use the system and performed at least one segmentation during the considered period (thus creating an entry in the server log files) were considered "active" users.

# Results

## Qualitative evaluation

In Fig. 3, we show how the "leg" model visually improved over the course of collaborative use, by showing the output of the automatic segmentation for two model snapshots: at initial pretraining and in November 2022. The proton-density contrast was acquired with a similar protocol as the dataset in which the initial models were pretrained; the T1 contrast is instead a new contrast that the initial pretrained model was never exposed to. A clear improvement can be observed for both image contrasts. The three representative slices were extracted from group B, as described above, and were not included in training of the model generations considered.

## Local validation

The average performance over all the datasets increased with each model generation (Fig. 4a), with an average increase of $0.009 \pm 0.001$ points (95% confidence interval 0.006-0.012) per generation on the group A datasets ($p < 0.001$). In absolute terms, the DSC increased from a median of 0.66 (interquartile range: 0.62–0.70) to 0.73 (interquartile range: 0.69–0.77).

A similar average DSC increase can be observed in group B (Fig. 4b). The linear model also had a linear coefficient of $0.007 \pm 0.002$ points (95% confidence interval 0.003-0.011) per generation ($p < 0.001$), thus demonstrating that the model incrementally trained on group A successfully generalized to the segmentation of group B. In absolute terms, the DSC increased from a median of 0.69 (interquartile range: 0.63–0.69) to 0.71 (interquartile range: 0.65–0.73).

## Usage statistics

Over the considered period, Dafne reached 36 active users, affiliated with institutions with MRI scanners from at least three major vendors (Siemens Healthineers, General Electric, and Philips Healthcare). The majority of users, according to the information shared by themselves in their initial access request, were students, either pursuing a PhD in muscle MRI, or a doctoral degree or specialization. The users collectively segmented 662 datasets. Of these, 639 were considered valid data points, and 23 (3.5%) points were excluded due to potential irregular user behavior, namely perfect dice scores ($>=0.99$, indicating that no refinement was performed) and dice scores very close to zero (<0.1, possibly indicating that a wrong segmentation task was performed).

The “thigh” model was used on 123 datasets, with a median DSC of 0.88 (interquartile range 0.82 – 0.91). The “leg” model was used on 516 datasets, with a median DSC of 0.82 (interquartile range 0.71 – 0.88).

Analysis of model evolution over time (Fig. 5) showed that the average DSC of the system declined as number of users increased, both for the leg and thigh datasets. This pattern was expected and is compatible with the introduction of different contrasts and protocols previously unknown to the original models; however, the DSC recovered over the course of the Dafne use and reached consistently high values after user number stabilization.

# Discussion

In this work, we demonstrated that Dafne, a fully open client-server software package for the segmentation of medical images, is able to incorporate new image contrasts in a controlled experiment involving the introduction of a new MRI contrast to the model and observing the model increases is performance with usage ($p < 0.001$). The real-world performance, measured by Dice similarity coefficients reported automatically by the client during usage, is able to adapt to the requirements of a growing set of users, by achieving overall improved quality over time, although potentially more slowly with respect to a typical training curve in centralized learning.

Specifically, the “leg” segmentation model showed an overall increase in its average performance in comparison with the early results. The “thigh” model reported high-performance values in its early stages, when the user pool was still small, followed by a decline and a subsequent recovery. This behavior is expected and is an indication of the initial low adaptation of the model to new image protocols and contrasts, which it incorporated in the later stages. These findings are consistent with those reported by Liang et al. (28), who demonstrated that models trained on homogeneous datasets require fine-tuning before being applicable to new protocols.

One key component of the task was developing a user-friendly interface for loading and manually segmenting (or refining the existing segmentation) medical datasets in multiple formats. This allowed for a natural workflow for medical professionals, who could proficiently exploit the system and have complete control over the end results. The feasibility and convenience of this proposed workflow was already demonstrated in a previous study by Wang et al (29).

Our continuously adapting approach offers an alternative to traditional model workflows, balancing performance improvements with necessary human oversight in clinical environments. While our data show overall improvement, performance fluctuations occur between model generations. For reproducibility needs, users can disconnect from Dafne and use a specific model version, later reconnecting to contribute their incrementally improved model — a feature that explains the reduced generation count in our validation study.

User-refined models undergo validation against representative server-stored data, including diverse cases such as muscles with and without fat replacement. This validation process may limit generalization for atypical cases (e.g., patients with implants), but our two-level validation approach intentionally prioritizes model stability and protection against potential degradation, ensuring reliability for the average user.

Drift or degradation in model performance may occur for certain data types or protocols that were previously learned but not represented in the validation data. Additionally, greater variability in performance on new data is also possible. These issues arise not only from the deep learning model itself but also from the preprocessing steps. In particular, bias field correction may yield inconsistent results across different MRI coil profiles or signal attenuations, thereby introducing an extra degree of input variability. Furthermore, the reliance on user refinement as a key part of the learning process may introduce biases based on individual user preferences, institutional practices, or varying levels of user expertise, potentially affecting the model's generalizability. This presents a drawback compared to a more conventional approach, where data are centralized and can be carefully checked and refined for accuracy. Delegating model refinement to inexperienced or unmotivated users may result in suboptimal learning. This issue is common in other federated learning scenarios, where data curation is not performed by a single entity, potentially leading to the inclusion of data of inconsistent quality.

A key advantage of our system is client-side incremental learning, which significantly reduces centralized computational requirements. Using optimized hyperparameters from (19), including a batch size of 5 and just 5 epochs, enables efficient processing on standard hardware—requiring only about 5 minutes on a CPU-only machine per dataset. Unlike conventional federated learning approaches that demand intensive site-wide training (10–14), our server operates on a simple CPU-only virtual machine, with validation being its only computationally significant task, made efficient by the AMSGrad algorithm's robust convergence properties (25).

While Dafne's privacy-preserving incremental approach is a strength, it also limits the ability to audit or trace the specific data contributions that led to model improvements, which is a potential concern in highly regulated medical environments. Additionally, while Dafne's client-side incremental learning reduces computational requirements, it may not be as effective as full model retraining for incorporating significant new features or adapting to dramatically different data distributions, potentially limiting its ability to evolve for radically new use cases without more substantial updates to the base model.

In this paper, we analyzed the two models that were initially offered with the Dafne package. However, recent development efforts have been devoted to creating a simple interface for the training of new deep learning models based on a small set of reference standard data. This interface, called “model trainer” is now able to produce a model file that can be directly imported into Dafne, which is capable of lifelong learning, with minimal user interaction. Through this tool, various other models (abdominal segmentation, kidney, spine) are deployed or in development, thus enabling Dafne to become a truly generic software for medical image segmentation. Currently, more than 100 users are registered on the Dafne server.

Other free software offer functionalities similar to Dafne, ranging from deep-learning algorithms, to GUI-based segmentation software, and federated learning programming frameworks. For a detailed comparison of available software, please see the Supplementary Material

In conclusion, Dafne's approach, which implements a lifelong learning approach to medical image segmentation enabled by human supervision, can improve and generalize with new data provided by the users. Despite potential limitations compared with more traditional centralized learning approaches, Dafne offers a viable and practical solution in real-world clinical situations, especially when not all data are not available simultaneously. The Dafne software package is an easily deployable, general-purpose tool for medical professionals and researchers alike to implement advanced deep learning algorithms in their workflows, without the need for programming knowledge or complex software setups.

## Acknowledgment

The Authors would like to acknowledge all the active users of Dafne, and their precious feedback. This work has been partly supported by the Swiss National Science Foundation (SNF) Spark grant number CRSK-3_196515. Part of the authors of this paper are members of the EURO-NMD European Reference Network.

# Figures

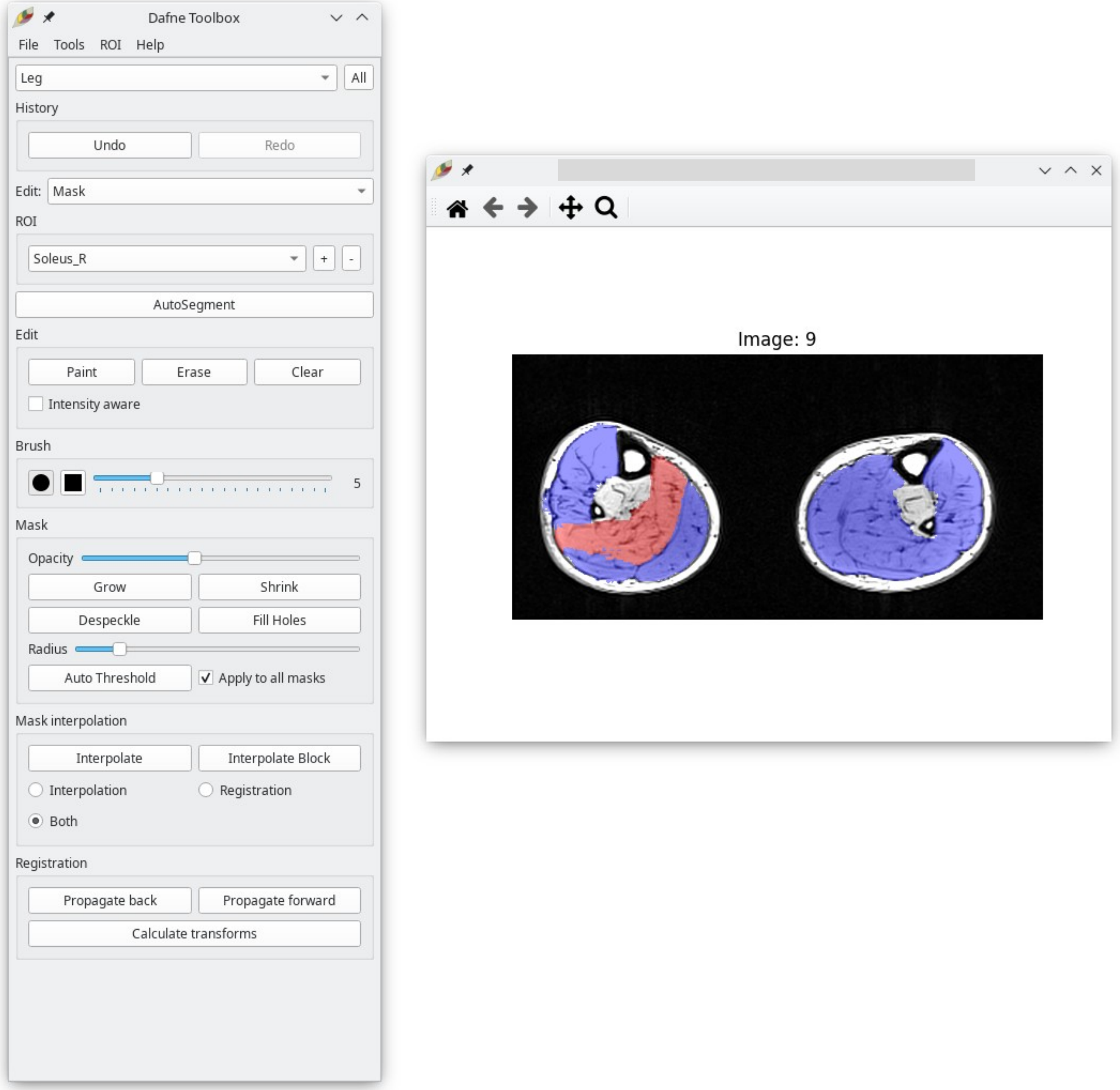

**Figure 1:** Screenshot of the Dafne user interface. Dafne is currently in "mask" edit mode and the corresponding tools are shown on the toolbox on the left. The active region of interest (in this case the right Soleus muscle) is overlaid in red in the image window. The inactive regions are overlaid in blue.

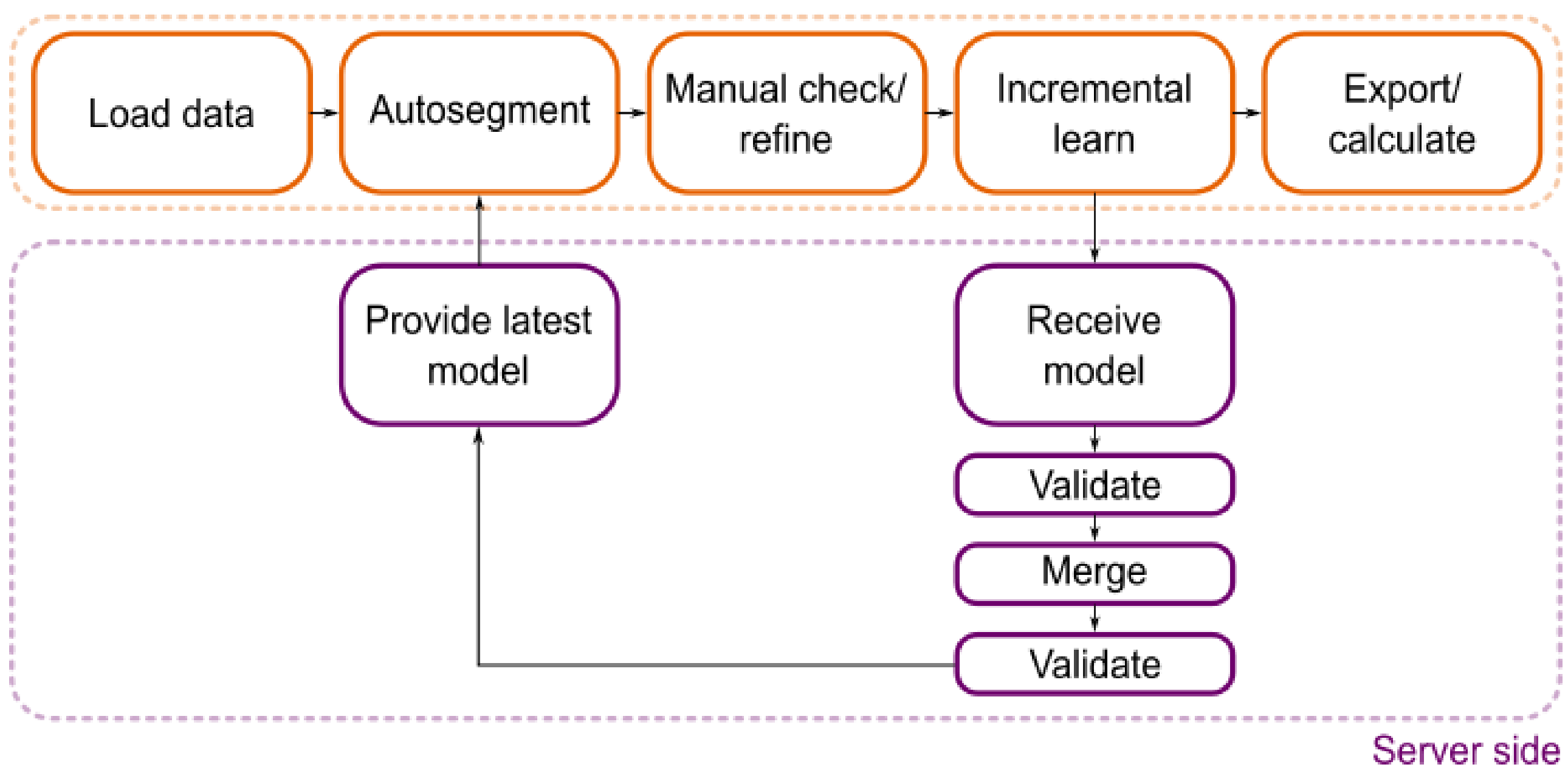

**Figure 2:** Dafne workflow. The top flow happens on the client's side, with user interaction in the manual refinement step. When the user exports the segmented masks, or calculates the voxel statistics from within the software, an incremental learning step over the used model is triggered.

The bottom flow happens automatically on the server once a refined model is received.

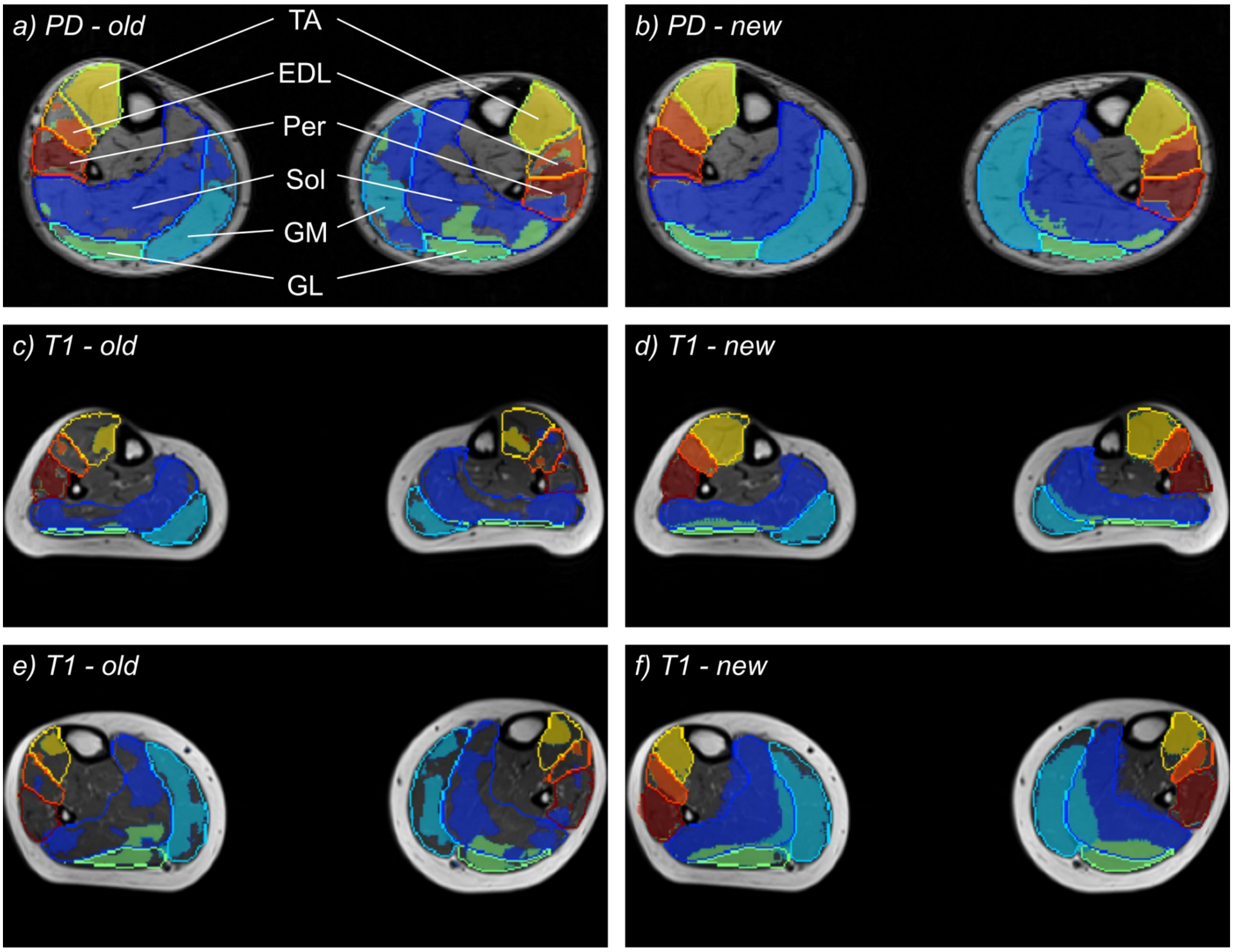

**Figure 3:** Segmentation example at two different time points and three different datasets. Panels a and b show a proton-density weighted dataset (the same contrast used in the pretraining) segmented with the initial pretrained model (a) and with the most current model (b). Panels c to f show T1-weighted datasets (a contrast not included in the pretraining) segmented with the initial model (c and e) and with the current model (d and f). The outlines around the muscles show a ground-truth manual segmentation, and the semitransparent areas show the result of the automatic segmentation. On the first panel, the labeling of the muscles is shown (TA: Tibialis Anterior; EDL: Extensor Digitorum Longus; Per: Peroneus; Sol: Soleus; GM: Gastrocnemius Medialis; GL: Gastrocnemius Lateralis). In the last panel, an overestimation of the gastrocnemii by the automatic algorithm is shown. This case is especially difficult because muscle boundaries are not clearly visible in the image.

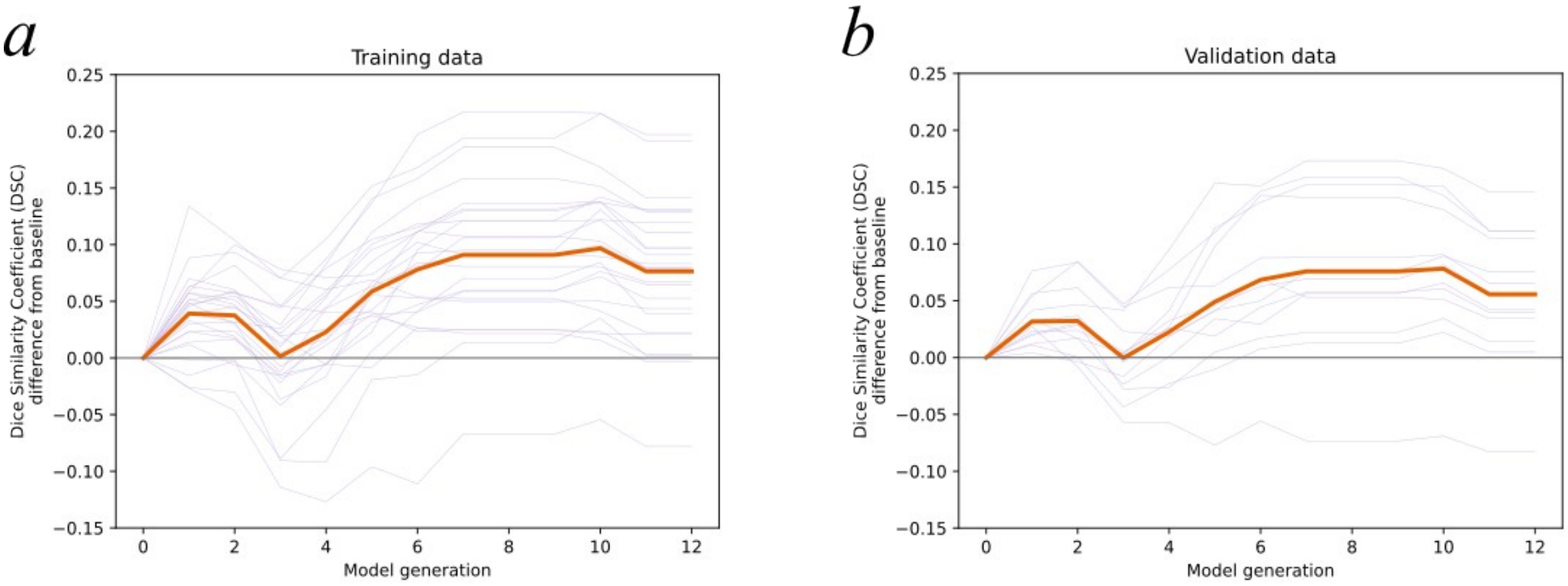

**Figure 4:** Evolution of the Dice Similarity Coefficients over the model versions produced by the local evaluation. Panel a reports the results for the data used for the training (group A in the text), whereas panel b reports the results of the data used for validation (group B). The thin gray lines correspond to each dataset, and the thick orange line is the average DSC across all datasets. Plots available under CC-BY at Ref. (26).

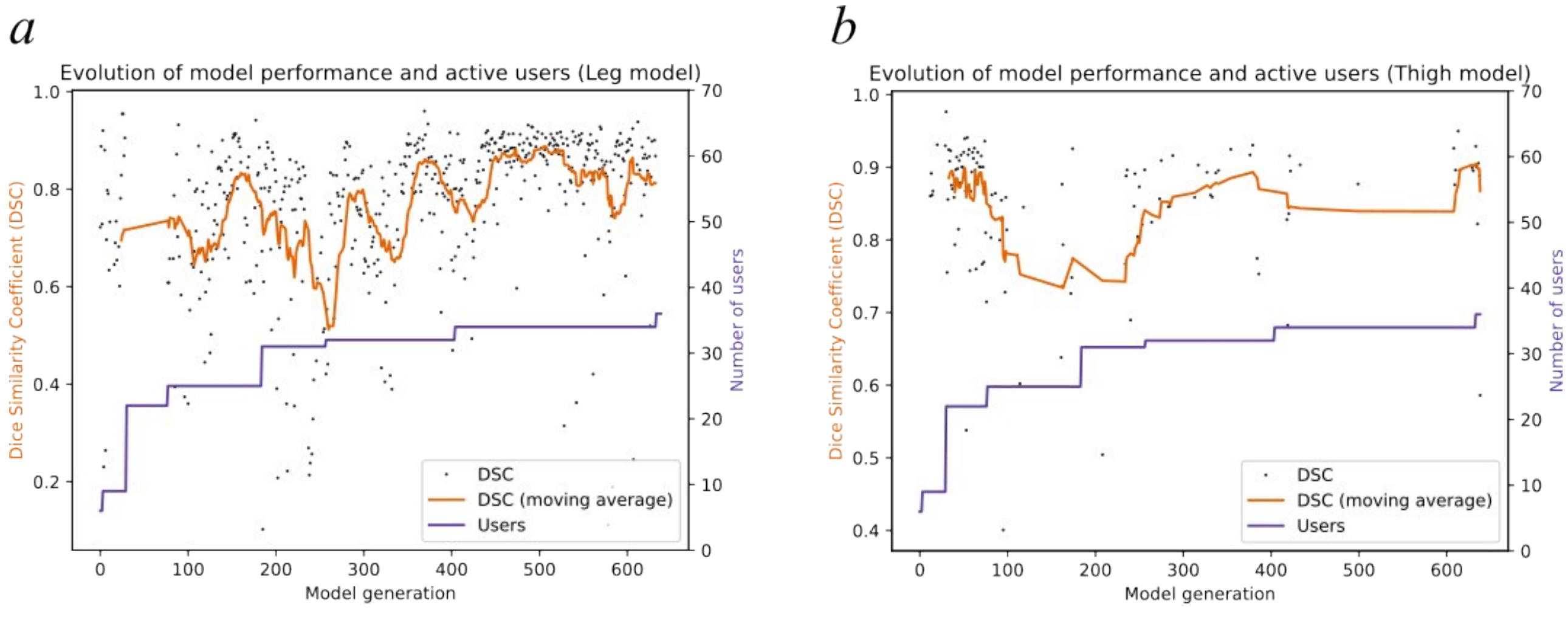

**Figure 5:** Evolution of model performance for both the leg (a) and thigh (b) models over time, as reported by the client, together with number of active users (in purple). For both models, an initial decline after user expansion can be observed, which then recovers after stabilization of user numbers. Plots available under CC-BY at Ref. (26).

# Supplementary Material: Comparison of software packages with similar functionality

Dafne's model architecture and training data are based on the work by Agosti et al. (1), although the software presented in this previous work does not contain the federated/continuous learning concepts introduced with Dafne. Popular GUI-based software packages that can be used for manual and automated organ and lesion segmentation are 3D Slicer (https://slicer.org/) and ITK Snap (https://itksnap.org) (2). ITK Snap is mostly focused on manual segmentation, providing some tools that are similar to Dafne such as edge detection and slice propagation; however, it does not contain advanced automated segmentation algorithms. Such advanced functionality can be accessed through the Distributed Segmentation Services (DSS), third-party-managed online services, which however require the sharing of the images with the service itself. 3D Slicer, on the other hand, has a more sophisticated plugin structure, which allows the local implementation of deep learning models, such as the recent and popular "Total Segmentator" model (3). Such plugins do not however natively support model updates based on user interaction, which is the core of Dafne functionality. On the other hand, other programming frameworks provide federated learning functionality, although they don't generally provide a readily established infrastructure, nor a graphical interface, and are intended to be included inside other software packages. Examples of these frameworks are NVIDIA Federated Learning Application Runtime Environment (FLARE, NVidia, Santa Clara, California, USA) (4) and its counterpart developed by Intel Open Federated Learning (OpenFL, Intel, Santa Clara, California, USA) (5). These are both general-purpose packages dedicated to the federated training of deep learning models, without particular focus on medical imaging or image segmentation. They provide programming interfaces and command-line tools for the deployment of servers and clients, and the distribution, training, and merging of models. They are mostly focused on more conventional federated learning approaches with respect to Dafne, in the sense that they are particularly suitable for situations where the training data are present at the same time at different locations. Lastly, a framework that is more focused on medical image analysis is the Medical Open Network for Artificial Intelligence (MONAI) (6), which is a set of Python libraries offering pretrained models, predefined model architectures, and, in general, functions that facilitate the implementation of software for medical image analysis. Similarly to the Dafne plugin structure, MONAI has the concept of "bundles", which are downloadable models that include not only the neural network, but also the required pre- and post-processing steps. In contrast with Dafne, however, MONAI does not natively offer a graphical interface, and is not primarily intended for continuous incremental learning.